\documentclass[english,aps,pra,twocolumn,showpacs,superscriptaddress]{revtex4}
\usepackage{amssymb}
\usepackage{amsmath}
\usepackage{graphicx}
\usepackage{subfigure}
\usepackage{amsfonts}
\usepackage{dcolumn}
\usepackage{bm}
\begin{document}

\title{The electronic Casimir-Polder force in a 1D tight-binding nanowire}

\author{Hui Yang}

\affiliation{School of Physics, Northeast Normal University, Changchun 130024,
People\textquoteright{}s Republic of China}

\author{Li-Ping Yang}

\affiliation{State Key Laboratory of Theoretical Physics, Institute of Theoretical
Physics and University of the Chinese Academy of Sciences, Beijing
100190, People's Republic of China}

\author{T. Y. Zheng}

\affiliation{School of Physics, Northeast Normal University, Changchun 130024,
People\textquoteright{}s Republic of China}
\begin{abstract}
We study the effect of two non-interacting impurity atoms near by
a one-dimensional nanowire, which is modeled as a tight-binding hopping
model. The virtual single-electron hopping between two impurities
will induce an additional energy depending on the distance of two
impurities, which gives a electronic Casimir-Polder effect. We find
that the Casimir-Polder force between the two impurities decreases
with the impurity-impurity distance exponentially. And the effects
of nanowire and finite temperature on the Casimir-Polder force are
also discussed in detail, respectively.
\end{abstract}

\pacs{31.30.jh, 73.22.-f, 81.07.Gf}

\maketitle

\section{introduction}

Environment-fluctuation-induced effects are ubiquitous in physics,
since a realistic system is inevitably coupled to an environment with
very large degrees of freedom. One important effect induced by the
fluctuation of the electromagnetic field (EMF) is the Casimir effect,
which was predicted by Casimir~\cite{Casimir} that there exists
an attractive force between two neutral, parallel and perfectly conducting
plates placed a few micrometers apart in vacuum. This attractive force
between two plates has been verified by numerous experiments~\cite{E1,E2,E3,E4,E5,E6,E7}.
In 1956, Lifshitz~\cite{Lifshitz} generalized the Casimir effect
induced by the quantum vacuum fluctuation to the case the attractive
force between two dielectric plates induced by the thermal fluctuation~\cite{Thermal_case,Thermal_case2,Thermal_case3}.
About ten years later, Boyer~\cite{Boyer} first discovered that
the Casimir force could be repulsive for a conducting spherical shell.
This set off a torrent of theoretical explorations~\cite{Lukosz,PRA98_Golestanian,PRL01,PRA05,PRL05}
on the geometric shape dependence of the Casimir force.

The Casimir force has been extensively applied in physics~\cite{PhysRep01},
especially in atomic physics. In 1948, Casimir and Polder~\cite{Casimir-Polder}
calculated the attractive force between two neutral polarizable atoms
(and the force between a neutral atom and a perfectly conducting wall)
in vacuum called Casimir-Polder (CP) force, which gives a significant
correction of the van der Waals-London force~\cite{London} for the
large atomic separation case. In most of the previous literatures~\cite{DW1,Mirror,BLHU,PRA07,BEC,TianT,HYang,YongLi},
researches were focused on the force induced by EMF fluctuation. Recently,
Tanaka \textit{et al}.~\cite{ECP} generalized to the case, where
the attractive force between two neutral impurity atoms results from
the exchange of virtual electrons, called electronic Casimir-Polder
(ECP) force.

In this paper, we study a realistic solid system composed of a one-dimensional
(1D) nanowire and two separate impurity atoms. Different from the
free-electron gas, the electron traveling in the nanowire described
by the tight-binding model has a cosine nonlinear dispersion relation,
which will present much more rich physics of interest. Due to the
finite energy-band width, the results without divergence are obtain.
The ECP force decreases exponentially when the distance between the
two charges increases. The decay rate of the ECP force is larger for
smaller hopping strength of the nanowire and it increases with the
absolute value of the detuning between the atom energy and the site
energy of the nanowire. For a fixed impurity-impurity distance, the
ECP force is larger for bigger hopping strength and it decreases with
the absolute value of the detuning. In the low-energy regime, the
cosine dispersion relation can be approximately expanded as a linear
quadratic one. When the energy of the impurities is close to the edge
of the energy band, we obtain the similar results as presented in
Ref.~\cite{ECP}. Additionally, unlike the well-known results for
the traditional CP force between two atoms, the ECP force obtained
in this system decreases with the temperature.

In the next section, the model Hamiltonian and the nonlinear dispersion
relation of the nanowire are presented. In Sec. III, we calculate
the electronic Casimir-Polder force between the two impurity atoms.
The numerical results are addressed in Sec. IV. In Sec. V, the thermal
ECP effect is taken into account. Finally, the summary of our main
results is given in Sec. VI.

\section{model setup and dispersion relation}

The system considered here is illustrated in Fig.~1. The one-dimensional
nanowire consists of $2N+1$ identical artificial atoms~\cite{Tight-Bindig_Model}
with the same site energy $\omega$ ($\hbar=1$). The electrons can
hop between neighboring atoms, where the hopping strength $J$ between
any two nearest-neighbor atoms is the same. The nanowire is described
by the typical tight-binding Hamiltonian,
\begin{equation}
H_{C}=\sum_{j=-N}^{N}[\omega c_{j}^{\dagger}c_{j}-J(c_{j}^{\dagger}c_{j+1}+c_{j+1}^{\dagger}c_{j})],
\end{equation}
where $c_{j}(c_{j}^{\dagger})$ is the electron annihilation(creation)
operator for the $j$th atom.

\begin{figure}
\includegraphics[width=8cm]{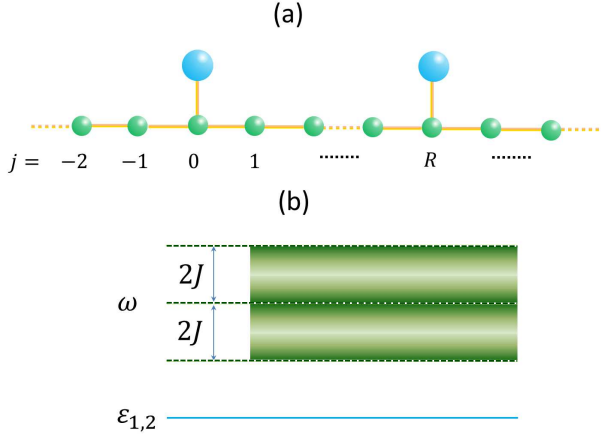}\\
\caption{\label{fig:schematic}(Color online) Schematic representation of the
model Hamiltonian. The two impurity atoms with energies $\varepsilon_{1}$and
$\varepsilon_{2}$ are placed at $x=0$ and $x=R$, respectively. }
\end{figure}

Two impurity atoms with energies $\varepsilon_{1}$ and $\varepsilon_{2}$
are placed on the nanowire at $x=0$ and $x=R$, respectively. The
impurities are weakly coupled to the nanowire, and the Hamiltonian
of the total system is split into two parts $H=H_{0}+H_{I}$ with
$H_{0}=H_{A}+H_{C}$. Here,

\begin{equation}
H_{A}=\varepsilon_{1}d_{1}^{\dagger}d_{1}+\varepsilon_{2}d_{2}^{\dagger}d_{2}
\end{equation}
is the Hamiltonian of impurities with $d_{i}(d_{i}^{\dagger})$ the
electron annihilation (creation) operator for the impurity states.
The interaction between the impurities and the nanowire is described
by
\begin{equation}
H_{I}=\lambda_{0}(c_{0}^{\dagger}d_{1}+d_{1}^{\dagger}c_{0})+\lambda_{R}(c_{R}^{\dagger}d_{2}+d_{2}^{\dagger}c_{R}),\label{eq:interaction}
\end{equation}
where $\lambda_{j}$ ($j=0,\ R$) is the corresponding coupling constant.

By using the Fourier transform,
\begin{equation}
c_{j}\!\!=\!\!\frac{1}{\sqrt{2N+1}}\!\sum_{k}\! c_{k}\! e^{ikj},\ c_{j}^{\dagger}=\!\!\frac{1}{\sqrt{2N+1}}\!\sum_{k}\! c_{k}^{\dagger}\! e^{-ikj}\!,
\end{equation}
$H_{C}$ can be diagonalized as
\begin{equation}
H_{C}=\sum_{k}\Omega_{k}c_{k}^{\dagger}c_{k},
\end{equation}
to obtain a nonlinear dispersion relation $\Omega_{k}=\omega-2J\cos k$.
As shown in Fig~1 (b), the eigenmodes of the nanowire form an energy
band symmetrically distributed around $\omega$ with width $4J$.
Additionally, the interaction Hamiltonian changes into
\begin{equation}
H_{I}=g_{1}\!\!\sum_{k}(c_{k}^{\dagger}d_{1}\!+\! d_{1}^{\dagger}c_{k})\!+\! g_{2}\!\!\sum_{k}(e^{-ikR}\! c_{k}^{\dagger}d_{2}\!+\! e^{ikR}\! d_{2}^{\dagger}c_{k})\!.
\end{equation}
Here, we have taken the lattice constant as unit, the atom-atom distance
$R$ is a positive integer, and $g_{1}=\lambda_{0}/\sqrt{2N+1}$ and
$g_{2}=\lambda_{R}/\sqrt{2N+1}$ are the regularized coupling strength.

\section{the electronic Casimir-Polder force}

Comparing to the conventional CP effect, the two impurity atoms play
the role of the neutral atoms and the non-excitated nanowire behaves
as the EMF with continuous modes. Due to the finite width of the energy
band of the nanowire, we obtain the exact ECP force without divergence.

For simplicity, all the calculations are done in the single-electron
subspace. Here, we assume that the energies of the impurities locate
below the energy band of the coupled chain, i.e., $\varepsilon_{1,2}<\omega-2J$.
The eigenstates of $H_{0}$ are as follows:
\begin{equation}
|1,0;0_{k}\rangle=d_{1}^{\dagger}|0,0;0_{k}\rangle,
\end{equation}
\begin{equation}
|0,1;0_{k}\rangle=d_{2}^{\dagger}|0,0;0_{k}\rangle,
\end{equation}
\begin{equation}
|0,0;1_{k}\rangle=c_{k}^{\dagger}|0,0;0_{k}\rangle,
\end{equation}
where $|0,0;0_{k}\rangle=|0,0\rangle\otimes|0_{k}\rangle$ is the
electron vacuum states of whole system.

In order to obtain the interaction between the two impurity atoms,
we make the Fr\"{o}hlich transformation
\begin{equation}
H_{eff}=e^{-S}He^{S}
\end{equation}
where
\begin{equation}
S\!=\!\sum_{k}\!\frac{g_{1}\!\!\left(\! d_{1}^{\dagger}c_{k}\!-\! c_{k}^{\dagger}d_{1}\!\right)}{\Omega_{k}-\varepsilon_{1}}\!+\!\frac{g_{2}\!\!\left(\! e^{ikR}\! d_{2}^{\dagger}c_{k}\!-\! e^{-ikR}\! c_{k}^{\dagger}d_{2}\!\right)}{\Omega_{k}-\varepsilon_{2}}\!,
\end{equation}
is an anti-Hermitian operator. When coupling strengths $\lambda_{0,R}$
satisfy $|\lambda_{0,R}|\ll|\Omega_{k}-\varepsilon_{1,2}|$, we can
obtain the effective Hamiltonian to the second order according to
the perturbation theory,
\begin{eqnarray}
H_{{\rm eff}} & \!\!= & \!\!\!\left(\!\!\varepsilon_{1}\!+\!\sum_{k}\!\frac{g_{1}^{2}}{\varepsilon_{1}-\Omega_{k}}\!\!\right)\!\! d_{1}^{\dagger}d_{1}\!+\!\!\left(\!\!\varepsilon_{2}\!+\!\sum_{k}\!\frac{g_{2}^{2}}{\varepsilon_{2}-\Omega_{k}}\!\!\right)\!\! d_{2}^{\dagger}d_{2}\nonumber \\
 &  & \!\!\!+\sum_{k}\left(\Omega_{k}+\frac{g_{1}^{2}}{\Omega_{k}-\varepsilon_{1}}+\frac{g_{2}^{2}}{\Omega_{k}-\varepsilon_{2}}\right)c_{k}^{\dagger}c_{k}\nonumber \\
 &  & \!\!\!\!+\!\sum_{k}\!\frac{g_{1}g_{2}e^{-ikR}}{2}\!\left(\!\!\frac{1}{\varepsilon_{1}-\Omega_{k}}\!+\!\frac{1}{\varepsilon_{2}-\Omega_{k}}\!\!\right)\! d_{1}^{\dagger}d_{2}\!\nonumber \\
 &  & \!\!\!\!+\!\sum_{k}\frac{g_{1}g_{2}e^{ikR}}{2}\!\left(\!\!\frac{1}{\varepsilon_{1}-\Omega_{k}}\!+\!\frac{1}{\varepsilon_{2}-\Omega_{k}}\!\!\right)\! d_{2}^{\dagger}d_{1}.
\end{eqnarray}

For convenience, we consider the symmetric case in which $\varepsilon_{1}=\varepsilon_{2}=\varepsilon_{0}$
and $\lambda_{0}=\lambda_{R}=\lambda$ (i.e., $g_{1}=g_{2}=g=\lambda/\sqrt{2N+1}$).
After diagonalizing the effective Hamiltonian $H_{eff}$, we can easily
get the eigenstates in the single-electron subspace,
\begin{equation}
|\varphi_{+}\rangle=\frac{1}{\sqrt{2}}(|1,0;0_{k}\rangle+|0,1;0_{k}\rangle),
\end{equation}
\begin{equation}
|\varphi_{-}\rangle=\frac{1}{\sqrt{2}}(|1,0;0_{k}\rangle-|0,1;0_{k}\rangle),
\end{equation}
\begin{equation}
|\varphi_{k}\rangle=|0,0;1_{k}\rangle,
\end{equation}
with eigenvalues
\begin{equation}
E_{+}=\varepsilon_{0}+\sum_{k}\frac{g^{2}(1+e^{-ikR})}{\varepsilon_{0}-\omega+2J\cos k},
\end{equation}
\begin{equation}
E_{-}=\varepsilon_{0}+\sum_{k}\frac{g^{2}(1-e^{-ikR})}{\varepsilon_{0}-\omega+2J\cos k},
\end{equation}
\begin{equation}
E_{k}=\left(\omega-2J\cos k\right)+\frac{2g^{2}}{\left(\omega-2J\cos k\right)-\varepsilon_{0}},
\end{equation}
respectively. As required in the former $\varepsilon_{0}-(\omega-2J)<0$,
the second order energy correction is negative, and then $|\varphi_{+}\rangle$
is the ground state of the system with $E_{+}$ the lowest-energy.
Changing the sum to integral with relation,
\[
\sum_{k}\rightarrow\frac{2N+1}{2\pi}\int_{-\pi}^{\pi}dk
\]
and omitting the $R$-independent term, we get the CP energy between
the two impurities by means of the residue theorem as
\begin{equation}
E_{cp}(R)=\frac{\lambda^{2}}{\Delta}\frac{1}{\sqrt{1-a^{2}}}\left(\frac{\sqrt{1-a^{2}}-1}{a}\right)^{R},
\end{equation}
where $\Delta=\varepsilon_{0}-\omega$ is the detuning between the
site energy of the nanowire and the impurity atom energy and $a=2J/\Delta\in(-1,0]$.

As the the lattice constant is taken as unit, we obtain the ECP force
between the two impurity atoms as
\begin{eqnarray}
f & \!\!\equiv & \!\!-[E_{cp}(R+1)-E_{cp}(R)]\nonumber \\
 & \!\!= & \!\!-\frac{\lambda^{2}}{\Delta}\frac{1}{\sqrt{1\!-\! a^{2}}}\!\!\left(\!\!\frac{\sqrt{1-a^{2}}\!-\!1}{a}\!\!\right)^{R}\!\![\!\frac{\sqrt{1\!-\! a^{2}}\!-\!1}{a}\!-\!1]\!,\label{eq:ECP_force}
\end{eqnarray}
Here, in the discrete lattice system, the ECP force was defined as
the difference of the CP energy. Obviously, the ECP force can be rewritten
as
\begin{equation}
f=-\frac{\lambda^{2}}{\Delta}\frac{1}{\sqrt{1-a^{2}}}[\frac{\sqrt{1-a^{2}}-1}{a}-1]e^{-\Gamma R}.
\end{equation}
where the decay rate $\Gamma=\ln[a/(\sqrt{1-a^{2}}-1)]$ and it increases
with $a$. It is ready to find that ${\rm lim}_{a\rightarrow0^{-}}\Gamma=\infty$,
and then the characteristic length $R_{C}=\Gamma^{-1}$ tends to $0$,
which means that there will be no Casimir-Polder force between the
two impurities if the hopping strength of the chain $J=0$.

\section{Discussion about the electronic Casimir-Polder force }

In the preceding section, we obtain the ECP force $f$, as given by
Eq.~(20), decreases with the atom-atom distance $R$ exponentially.
It is also dependent on the hopping strength $J$, the detuning $\Delta$,
and proportional to the square of the impurity-nannowire coupling
constant $\lambda^{2}$.

\begin{figure}
\includegraphics[width=8cm]{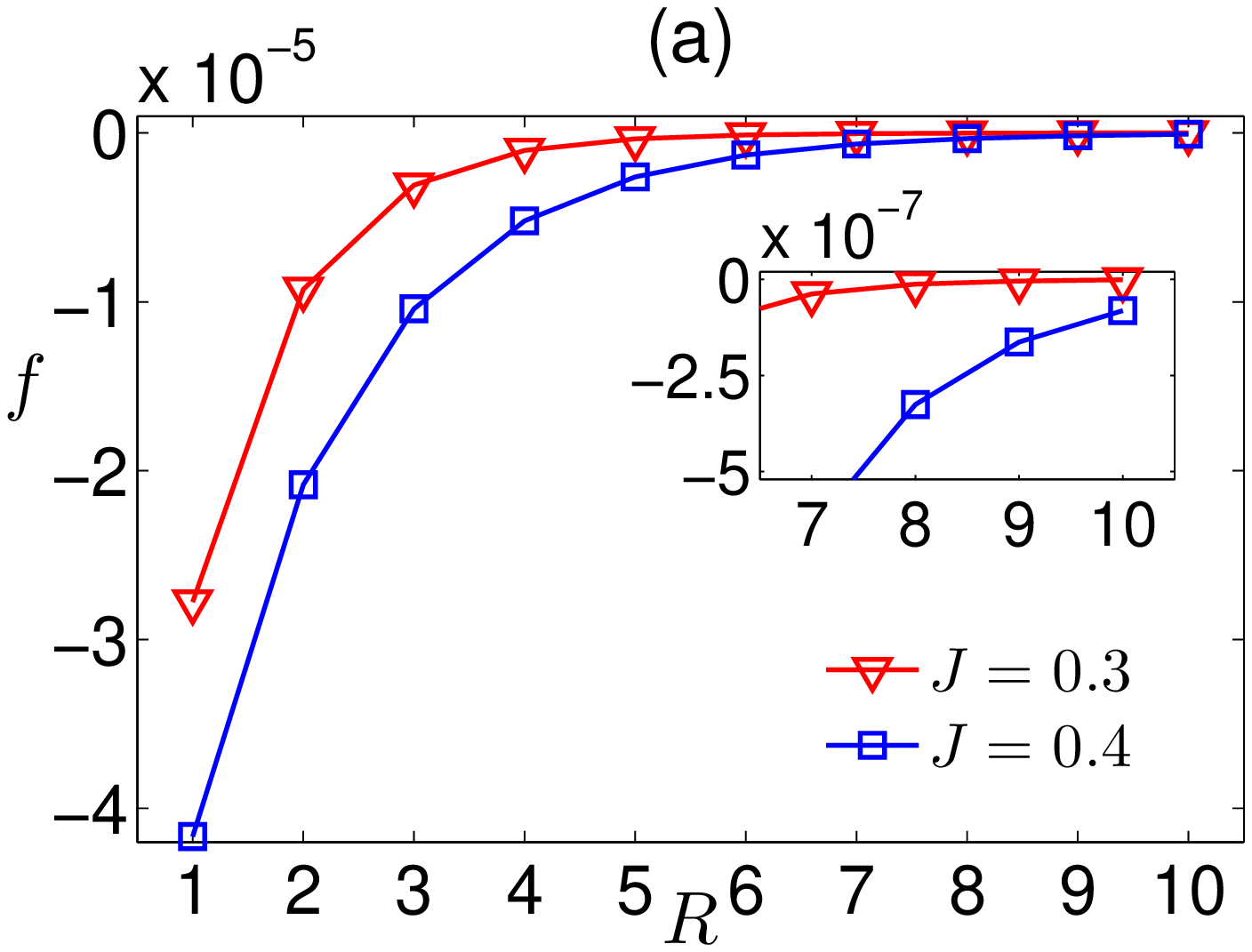}

\includegraphics[width=8cm]{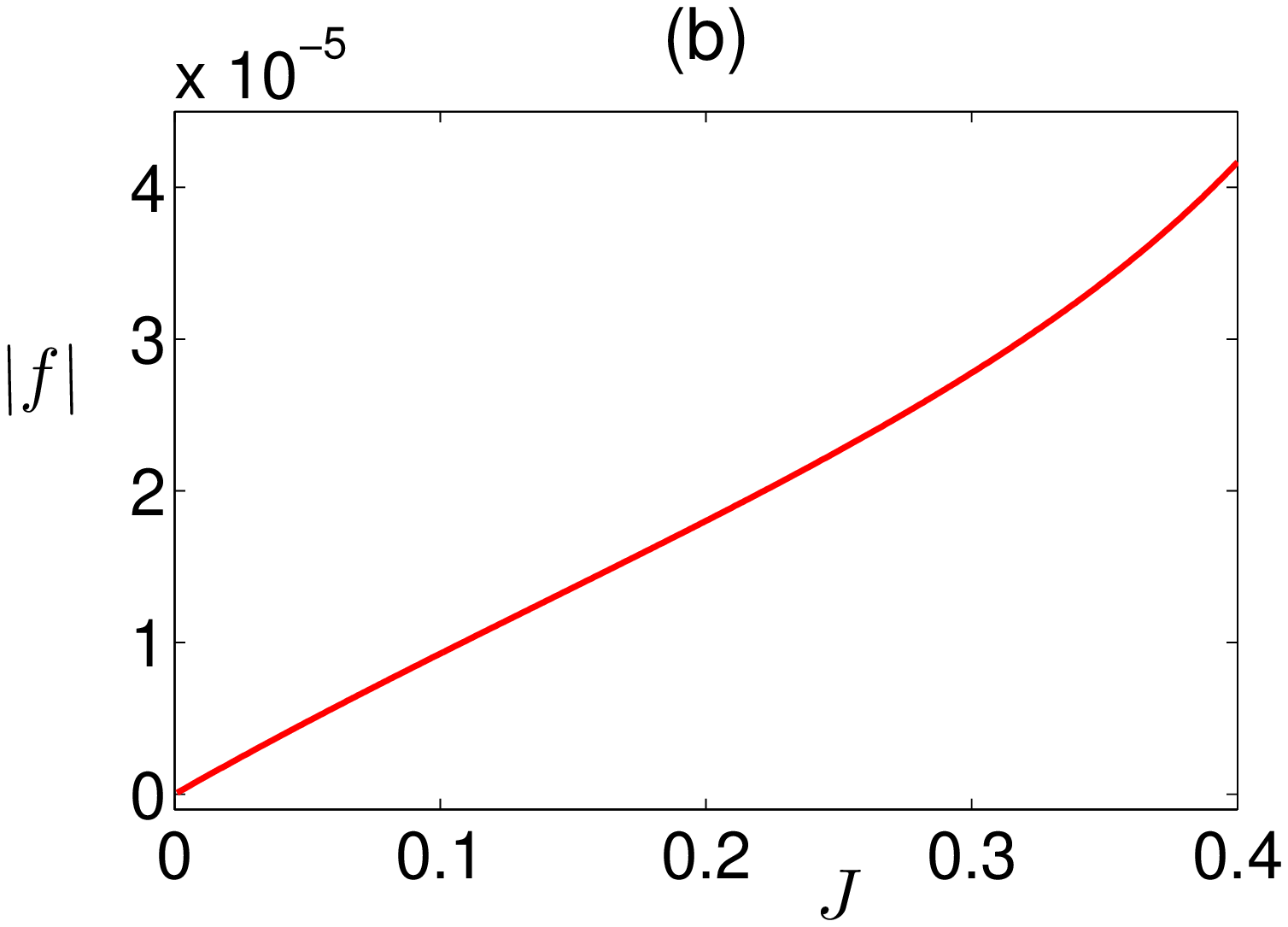}

\caption{(Color online) Here, we have taken the lattice constant and the impurity
energy as unit $\varepsilon=1$. And the other parameters are chosen
as $\lambda=0.01$ and $\varepsilon_{0}-\omega=-1$. (a) The electronic
Casimir-Polder forces $f$ vs. the atom-atom distance $R$ for different
values of $J$ are presented. The red-triangle line and blue-square
line represent the case of $J=0.3$ and $J=0.4$, respectively. (b)
The Casimir-Polder force $f$ vs hopping energy $J$ with fixed atom-atom
distance $R=1$.}
\end{figure}

For $J=0$, the CP energy is $E_{cp}(R)=0$ and $\Gamma\rightarrow\infty$
as shown in the former section. That is, there is no interaction between
the two impurities when the electrons cannot hop between neighboring
atoms. For $J\neq0$, we consider different values of $J$ to investigate
its influence on the force. The variation of the Casimir-Polder force
between the two impurity atoms with the atom-atom distance $R$ for
different values of $J$ is shown in Fig.2(a). The ECP force with
$J=0.3$ is represented by the red-triangle line and the one with
$J=0.4$ is depicted by the blue-square line. As shown in Fig~2(b),
for a fixed value of $R$, the force increases with the hopping strength
$J$.

In Fig~.3(a), we give the variation of the ECP force with $R$ for
different detuning $\Delta=\varepsilon_{0}-\omega$. The red-triangle
line and the blue-square line depict the cases $\Delta=-2$ and $\Delta=-3$,
respectively. As shown in Fig.~3(b), the ECP force decreases with
the absolute value of the detuning $\Delta$.

\begin{figure}
\includegraphics[width=8cm]{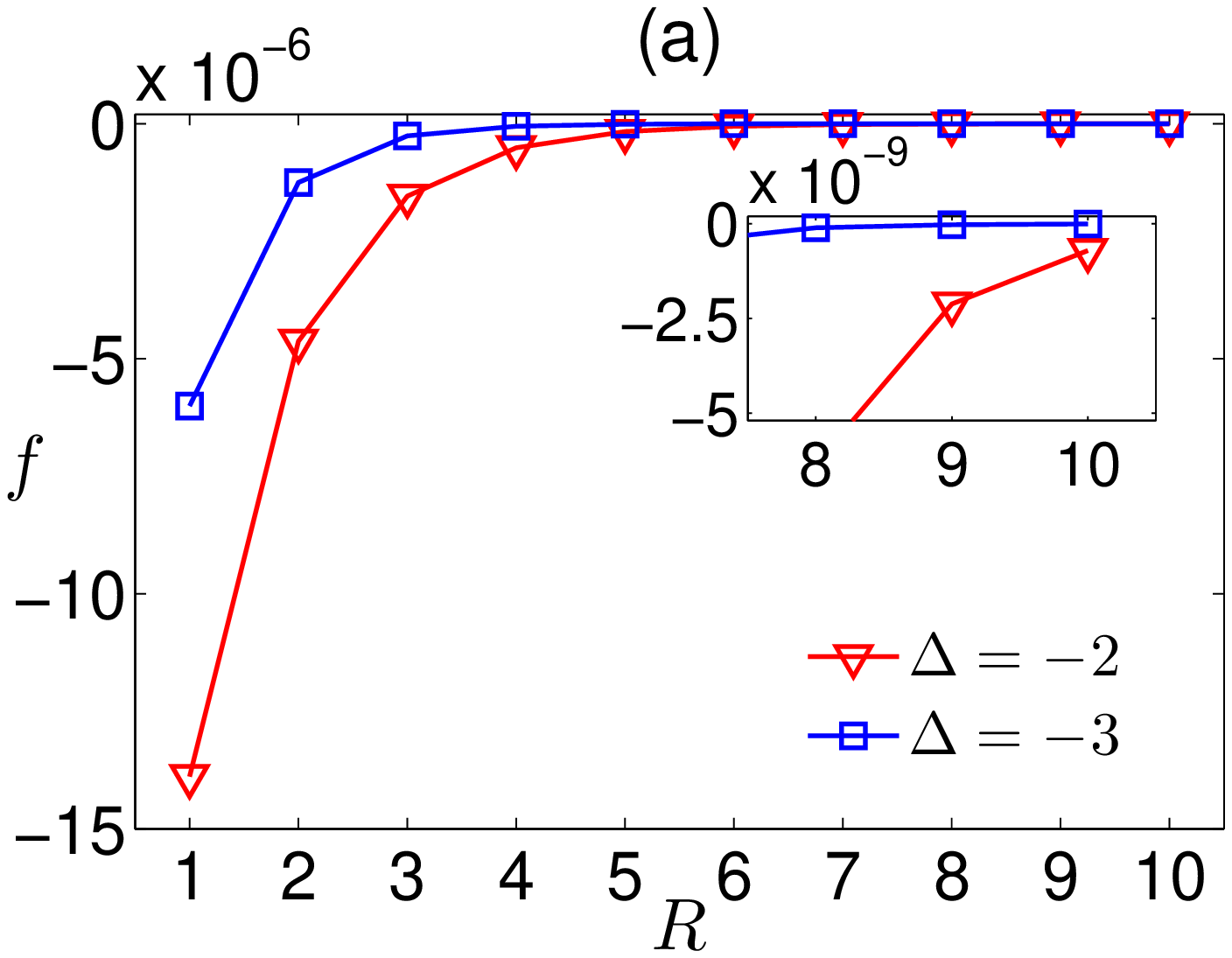}

\includegraphics[width=8cm]{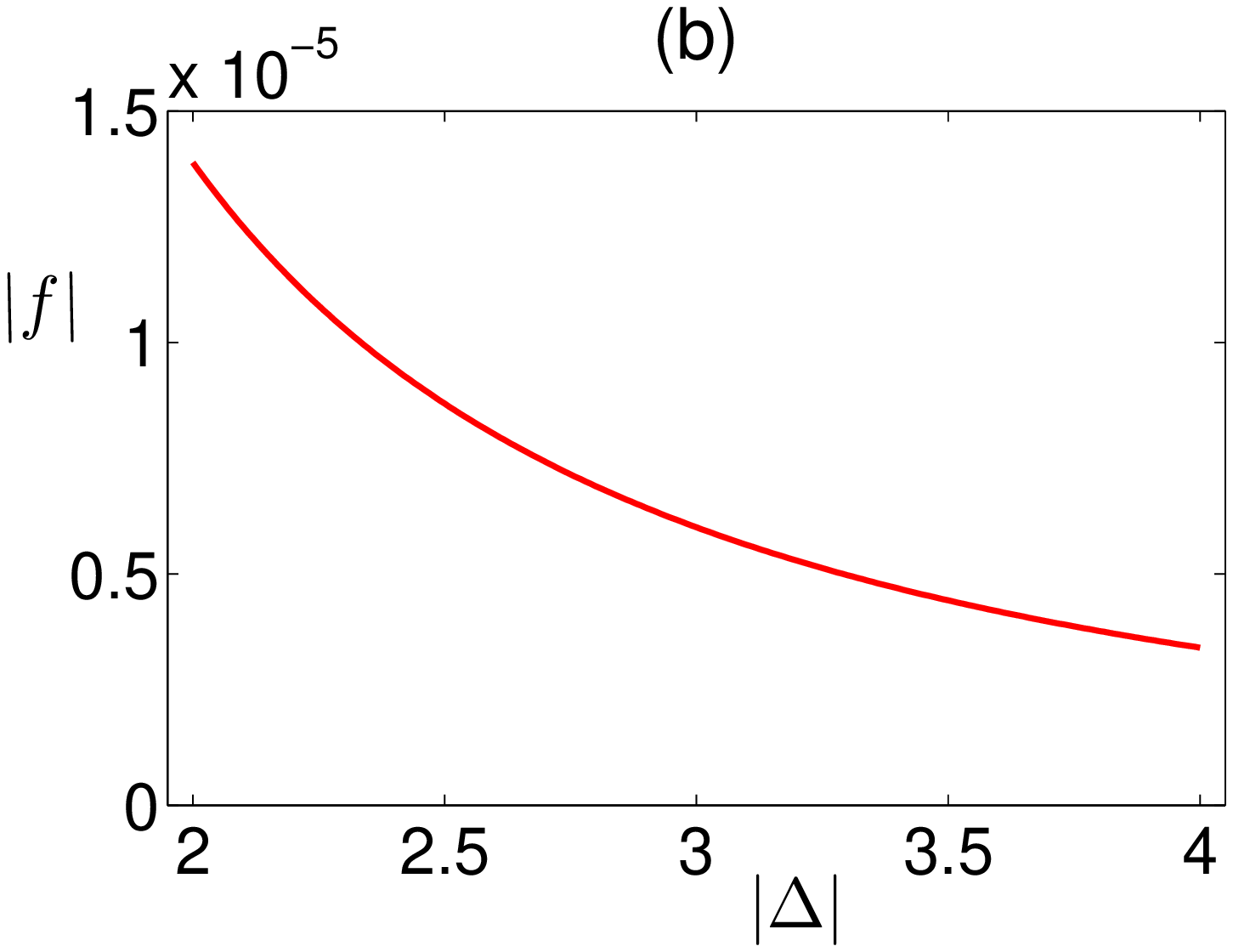}

\caption{(Color online) Here, we have taken the lattice constant and the impurity
energy as unit $\varepsilon=1$. And the other parameters are chosen
as $\lambda=0.01$ and $J=0.6$. (a) The electronic Casimir-Polder
forces $f$ vs the atom-atom distance $R$ for different values of
$\Delta$ are presented. The red-triangle line and blue-square line
represent the cases of $\Delta=-2$ and $\Delta=-3$, respectively.
(b) The Casimir-Polder force $f$ vs the detuning $\Delta$ with fixed
atom-atom distance $R=1$. }
\end{figure}

As shown in the former section, the ECP force between the two impurity
atoms decreases with the atom-atom distance exponentially $f\propto\exp(-\Gamma R)$.
The decay rate is determined by the parameter $J$ , $\varepsilon_{0}$,
and $\omega$, more specifically, $\Gamma$ is determined by $a=2J/\Delta\in(-1,0]$.
In Fig.~4, we find that decay rate $\Gamma$ increases with $a$
monotonically and tend to $\infty$ at the point $a=0$. That is,
the force decays slower with the impurity-impurity distance for larger
hopping strength and decays faster for larger absolute value of the
detuning. When the hopping strength $J\rightarrow0$, the ECP force
between the two atoms disappears.

\begin{figure}
\includegraphics[width=8cm]{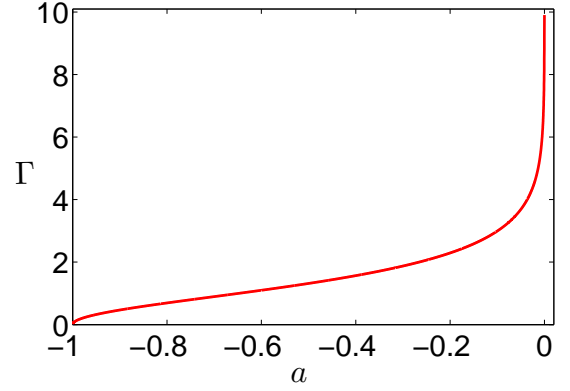}\caption{(Color online) The variation of decay rate $\Gamma$ with $a$.}
\end{figure}

In the low-energy regime $k\simeq0$, the nonlinear dispersion relation
$\Omega_{k}=\omega-2J\cos k$ can be be rewritten as $\Omega_{k}\simeq\omega-2J+Jk^{2}$,
by expanding the cosine function as $\cos k\simeq1-k^{2}/2$ approximately.
Then the Casimir-Polder energy is approximated as
\begin{equation}
E_{cp}(R)=\frac{2N+1}{2\pi}\int_{-\pi}^{\pi}dk\frac{g^{2}e^{-ikR}}{\varepsilon_{0}-\left(\omega-2J+Jk^{2}\right)}.
\end{equation}
If the difference of the atom energy and the lower edge of the energy
band is much smaller than the width of the energy band $\varepsilon_{0}-(\omega-2J)\ll4J$,
the integral over $k$ can be extended to $(-\infty,\,\infty)$. Then
the Casimir-Polder energy is given by
\begin{equation}
E_{CP}=-\frac{\lambda^{2}}{2Jb}e^{-bR},
\end{equation}
where $b=\sqrt{(\omega-2J-\varepsilon_{0})/J}$. The similar results
are obtained in Ref.~\cite{ECP}, where a quadratic dispersion relation
was used.

\section{The electronic Casimir-Polder force at finite temperature}

In the former sections, we only considered the zero temperature case.
Here, we will study the influence of the temperature on the ECP force.
At finite temperature $T$, the density matrix of the the system can
be given as
\begin{equation}
\rho_{T}\!=\!\frac{1}{Z}\left(\!\sum_{\alpha=\sigma}\! e^{-\beta E_{\sigma}}|\varphi_{\sigma}\!\rangle\!\langle\!\varphi_{\sigma}|\!+\!\sum_{k}\! e^{-\beta E_{k}}\!|\varphi_{k}\rangle\!\langle\varphi_{k}|\!\right)\!,
\end{equation}
where $Z=\sum_{\sigma=\pm}\exp(-\beta E_{\sigma})+\sum_{k}\exp(-\beta E_{k})$
and $\beta=1/k_{B}T$ is the inverse temperature. After integrating
over $k$, we rewrite the eigenvalues of the effective Hamiltonian
$H_{eff}$ in the single-electron space as

\begin{equation}
E_{+}\!\!=\!\varepsilon_{0}+\frac{\lambda^{2}}{\Delta}\frac{1}{\sqrt{1\!-\! a^{2}}}\!\![1+\left(\!\!\frac{\sqrt{1-a^{2}}\!-\!1}{a}\!\!\right)^{R}]\!,\label{eq:E_plus}
\end{equation}
and
\begin{equation}
E_{-}\!\!=\!\varepsilon_{0}+\frac{\lambda^{2}}{\Delta}\frac{1}{\sqrt{1\!-\! a^{2}}}\!\![1-\left(\!\!\frac{\sqrt{1-a^{2}}\!-\!1}{a}\!\!\right)^{R}]\!,\label{eq:E_negative}
\end{equation}
Then we obtain the average energy of the system $E_{T}={\rm Tr}(\rho_{T}H_{{\rm eff}})$
at temperature $T$ as
\begin{equation}
E_{T}=\frac{1}{Z}\{E_{+}e^{-\beta E_{+}}+E_{-}e^{-\beta E_{-}}+\sum_{k}E_{k}e^{-\beta E_{k}}\}.
\end{equation}
The thermal ECP force is defined as
\begin{equation}
f_{T}=-[E_{T}(R+1)-E_{T}(R)].\label{eq:ECP_force_T}
\end{equation}

Evidently, in the zero-temperature limit, the thermal ECP~(\ref{eq:ECP_force_T})
returns to~(\ref{eq:ECP_force}) obtain in Sec.~III. We plot the
variation of the ECP force $f$ with $R$ for different values of
$T/\varepsilon$ to investigate the influence of temperature on the
ECP force in Fig.~5. The values of $T/\varepsilon$ represented by
the red-triangle line, blue-square and green-star line are $0$, $0.1$,
and $1$, respectively. We see clearly that for a fixed value of $R$,
the ECP force decreases with the temperature $T$. As is seen in Eqs.~(\ref{eq:E_plus})
and (\ref{eq:E_negative}), the $R$-dependent potential induced by
the nanowire is attractive when the system is in the state $|\varphi_{+}\rangle$,
while the potential is repulsive when the system is in the state $|\varphi_{-}\rangle$.
So the ECP force is weaker at nonzero temperature than at zero temperature.
When the temperature increases, the probability of the system in the
state $|\varphi_{-}\rangle$ becomes larger and the probability in
the state $|\varphi_{+}\rangle$ decreases, so the ECP force decreases
with temperature. Obviously, this result is different with the well-known
results for the Casimir-Polder force between two atoms at finite temperature~\cite{temperature}.

\begin{figure}
\includegraphics[width=8cm]{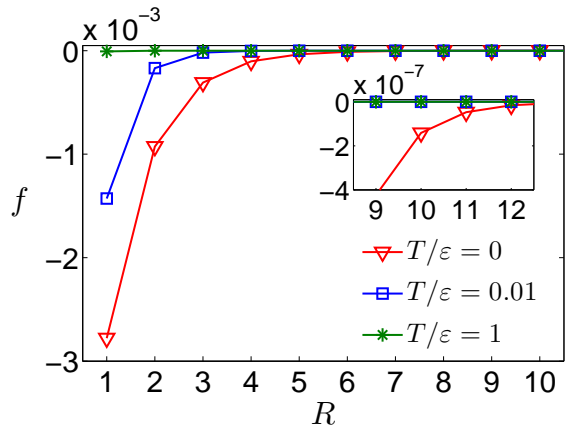}

\caption{(Color online)The Casimir-Polder force $f$ vs the atom-atom distance
$R$ for different values of $T/\varepsilon$ are presented. Here,
we have taken the lattice constant and the impurity energy as unit
$\varepsilon=1$. And the other parameters are chosen as $\lambda=0.1$,
$\varepsilon_{0}-\omega=-1$ and $J=0.3$. The red-triangle line,
blue-square and green-star line represent the values for $T/\varepsilon=0$,
$T/\varepsilon=0.1$ and $T/\varepsilon=1$, respectively. }
\end{figure}

\section{Conclusion}

We studied the ECP force between two non-interacting impurity atoms
near by a one-dimensional nanowire. In the single-electron space,
this ECP force the moving electron in the nanowire induced is attractive.
Different from the free-electron gas, the electron traveling in the
nanowire described by the tight-binding model has a nonlinear dispersion
relation. Based on the perturbation theory, we obtained the analytical
expression of the force without any divergence.

Our results show that the force between the two impurities decreases
with the impurity-impurity distance exponentially. The decay rate
decreases when the hopping strength of the nanowire increases, while
it becomes larger when the absolute value of the detuning between
the atom energy and the site energy of the nanowire increases. When
the hopping strength equals to zero, the decay rate becomes infinite
large and then the force between the two atoms disappears. We find
that the force induced by the nanowire also increases with the hopping
strength and decreases with the absolute value of the detuning. Based
on this result, we may use the nanowire to control the CP force in
practical application.

Besides, we also studied the ECP force at finite temperature. We find
that the force at nonzero temperature is weaker than that at zero
temperature and it decreases with the temperature, which is different
with the well-known results for the traditional CP force between two
atoms at finite temperature~\cite{temperature}. We wish this phenomenon
could be verified by experiments.
\begin{acknowledgments}
We thank Prof. C. P. Sun for helpful discussion. \end{acknowledgments}

\end{document}